        [1999/12/01 v1.4c Il Nuovo Cimento]
\documentclass{cimento}
\usepackage{graphicx,epsfig}

\title{Statistical analysis of RHESSI GRB database}
\shorttitle{Statistical analysis of RHESSI GRB database}
\shortauthor{J.~\v{R}\'{i}pa, R. Hudec, A.~M\'{e}sz\'{a}ros, ETC.}

\author{J.~\v{R}\'{i}pa\from{MFF}\from{AsU},
R.~Hudec\from{AsU}\thanks{\texttt{rhudec@asu.cas.cz}},
A.~M\'{e}sz\'{a}ros\from{MFF},\\ W.~Hajdas\from{PSI} \atque
C.~Wigger\from{PSI}}

\instlist{\inst{MFF} Astronomical Institute, Faculty of
Mathematics and Physics,\\Charles University~- Prague, Czech
Republic \inst{AsU} Astronomical Institute, Academy of Sciences of the~Czech Republic~-\\
Ond\v{r}ejov, Czech Republic \inst{PSI} Paul Scherrer Institut~-
Villigen, Switzerland}

\PACSes{\PACSit{98.70.Rz}{$\gamma$-ray sources; $\gamma$-ray
bursts}}

\begin{document}

\maketitle

\begin{abstract}
The Gamma-ray burst (GRB) database based on the data by the RHESSI
satellite provides a unique and homogeneous database for future
analyses. Here we present preliminary results on the duration and
hardness ratio distributions for a sample of 228 GRBs observed
with RHESSI.
\end{abstract}

\section{Introduction}

The Ramaty High-Energy Solar Spectroscopic Imager (RHESSI) is
a~NASA Small Explorer satellite designed to study hard X-rays and
gamma-rays from solar flares~\cite{ref:1}. It~consists mainly of
an imaging tube and a~spectrometer. The spectrometer consists of
nine germanium detectors~\cite{ref:2}. They are only lightly
shielded, thus making RHESSI also very useful to detect non-solar
photons from any direction. The energy range sensitive for GRB
detection extends from about 50~keV up to 20~MeV depending on the
incoming direction. Energy and time resolutions are excellent for
time resolved spectroscopy: $\Delta E$~=~3~keV (at 1000 keV),
$t$~=~1~$\mu$s. The effective area for near axis direction of
incoming photons reaches up to 200~cm$^2$ at 200~keV. With a field
of view of about half of the sky, RHESSI observes about one
gamma-ray burst per week (see also
\texttt{http://grb.web.psi.ch}).

\section{Duration distribution, hardness ratio vs. duration}

Totally observed 228 gamma-ray bursts from Feb. 2002 to Jan. 2006
were used. Originally it was found (results from BATSE, Konus-Wind
etc. instruments~\cite{ref:3,ref:4}) that there exist at least two
subclasses of GRBs; the short one with $T_{90}$ approximately less
than 2~s and the long one with $T_{90}$ approximately more than
2~s, where $T_{90}$ is the time interval during which the
cumulative counts increase from 5\% to 95\% above background. As
one can see, we obtained distribution with two maxima: about 0.2~s
and 15~s. Some articles point to existence of the three subclasses
of GRBs~\cite{ref:5}. We have investigated this in the RHESSI
database. We fitted one-, two- and three-lognormal functions
(fig.~1) and then we used statistical $\chi^2$-test to evaluate
these fits. This procedure was successfully used on BATSE Catalog
(see~\cite{ref:6}). From the~statistical point of view a~single
lognormal does not fit the~observed distribution. The~assumption
that there is only one subclass is rejected on a~greater than
99.99\% significance level. For the~best fitting with the~two
Gaussian functions we obtained significance level~=~15.9\%. For
the~best fitting with the~three Gaussian functions we obtained
significance level~=~16.4\%. Hence, RHESSI data can be interpreted
by at least two GRB's subgroups. A~further result is that the
short bursts give about 15\% of all 228 events observed by RHESSI.

\begin{figure}[th!]
\centerline{\epsfxsize=135mm \epsfbox[1 1 407
150]{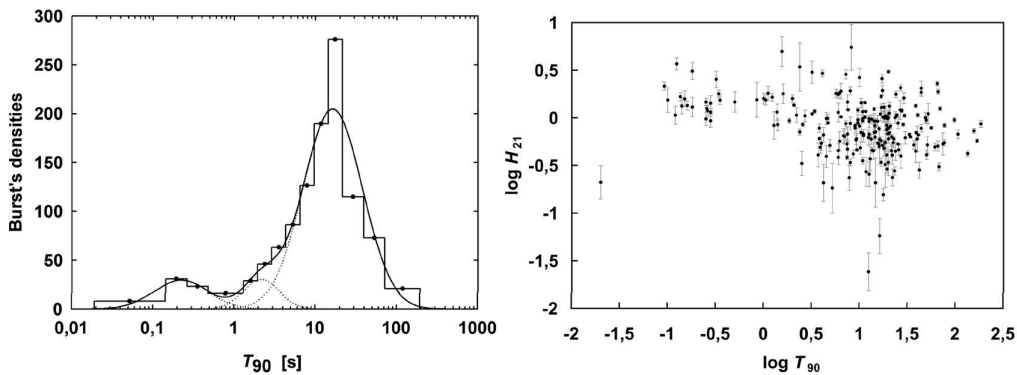}} \caption{(Left)~Distribution of durations of
GRBs with the best 3-lognormal fit. (Right)~Hardness ratio
$H_{21}$ vs. duration $T_{90}$. }
\end{figure}

The hardness ratio is defined as~the ratio of two fluences $F$ in
two different energy bands integrated over the~time interval
$T_{90}$. Specifically we have three energy bands: 25~-~120~keV,
120~-~400~keV and 400~-~1500~keV, and corresponding fluences
therein: $F_1$, $F_2$ and $F_3$. In fig.~1 we show
the~hardness-ratio $H_{21} = F_2/F_1$ vs. $T_{90}$ duration for
our RHESSI dataset. We confirm the~results of BATSE (\cite{ref:3})
that on average short GRBs are harder. We plan to use some cluster
analysis to exactly decide the~number of subgroups.

\acknowledgments We acknowledge the support by the
GA~AS~CR~3003206, OTKA grant T48870 and partly ESA PECS98023.

\end{document}